\documentclass[aps,prl,reprint,groupedaddress,floatfix]{revtex4-1}
\usepackage{graphicx,amsmath,amssymb,amsfonts,mathtools,booktabs,array}
\usepackage[utf8]{inputenc}
\usepackage[T1]{fontenc}
\usepackage{lmodern}
\usepackage[colorlinks=true,linkcolor=blue,citecolor=blue,urlcolor=blue,breaklinks=true]{hyperref}
\usepackage{microtype}
\DisableLigatures[f]{encoding=*,family=*}
\newcommand{\E}{\mathbb E}
\newcommand{\Var}{\operatorname{Var}}
\newcommand{\Cov}{\operatorname{Cov}}

\newcommand{\ii}{\mathrm i}
\newcommand{\dd}{\,\mathrm d}
\newcommand{\Rnoise}{\mathcal R}

\begin{document}
\title{Transmission and filtering of biological rhythms in stochastic gene expression}
\author{Chen Jia$^{1,2}$}
\email{Contact author: chenjia@hit.edu.cn}
\affiliation{$^1$ School of Mathematics, Harbin Institute of Technology, Harbin 150001, China}
\affiliation{$^2$ Beijing Research Institute, Harbin Institute of Technology, Beijing 100144, China}
% Author information retained from the supplied template; editable.

\begin{abstract}
% To be written after the main text has been revised.
How biological rhythms in upstream regulatory signals are transmitted to downstream stochastic gene expression is a fundamental question in systems biology. Although experiments have shown that molecular degradation strongly influences downstream rhythmic responses, a quantitative theory of rhythm transmission through bursty gene expression remains lacking. Here we develop a general analytical framework based on a martingale-based orthogonal decomposition approach, applicable to arbitrarily complex upstream dynamics. We derive exact autocorrelation and power-spectrum transmission laws, establish an oscillation-efficiency contraction principle, and identify a critical degradation rate required for rhythm transmission. An explicit transmission relation reveals how molecular degradation, burst frequency, modulation depth, and upstream timing variability jointly govern rhythm transmission and filtering. In particular, increasing burst frequency facilitates transmission, whereas uniformly rescaling burst sizes at fixed burst frequency leaves oscillation efficiency unchanged. Applied to the p53 regulatory system, the predicted degradation threshold is consistent with the experimentally observed distinction between pulsing and nonpulsing responses across ten target proteins.
\end{abstract}

\maketitle

% ===== Begin introduction =====
\section{Introduction}
Gene expression in living cells is frequently driven by time-varying upstream regulatory signals. Cell-cycle progression and circadian clocks regulate the temporal patterns of transcription, while pulsatile regulators such as p53 and NF-$\kappa$B coordinate cellular stress responses \cite{nicolas2018modulation, skinner2016single, lahav2004dynamics, nelson2004oscillations}. Importantly, biological regulatory signals carry temporal structure beyond their average levels, including pulse frequencies, durations, and temporal correlations. How this temporal structure is processed and transmitted through downstream gene expression networks is therefore a fundamental question in quantitative biology. In this process, gene expression systems act as stochastic filters: molecular degradation determines how rapidly downstream gene products respond to upstream fluctuations, while stochastic bursting introduces intrinsic noise that can obscure rhythmic signals. Experimental studies have shown that short-lived gene products can closely track upstream rhythms, whereas long-lived products tend to integrate successive pulses, thereby attenuating or filtering out upstream rhythms \cite{robertson2008real, cesbron2013light, porter2016p53, hanson2019protein, zambrano2016nf}. In particular, targets of the same transcription factor can exhibit markedly different temporal responses depending on their gene product stability \cite{porter2016p53, hanson2019protein}. These observations raise a fundamental question: under what conditions can stochastic rhythms in upstream regulatory signals be transmitted to downstream gene expression?

Previous studies \cite{song2019frequency, gupta2022frequency} have established exact relations connecting the power spectra of upstream reaction rates to those of downstream molecular fluctuations, laying a foundation for understanding stochastic signal propagation in gene expression. General quantitative relations between upstream signal fluctuations and downstream gene expression noise have also been recently derived \cite{mccune2026transcription}. However, these studies have not explicitly addressed how temporal structures associated with upstream rhythms are transmitted through bursty gene expression, particularly when both burst frequency and burst size are modulated by upstream signals. In particular, a general quantitative criterion determining whether downstream gene expression preserves or filters out upstream rhythmic fluctuations, and how this process depends jointly on stochastic bursting and molecular degradation, remains lacking.

In this study, we develop a general analytical framework for understanding how biological rhythms are transmitted through stochastic gene expression. Using a novel martingale-based orthogonal decomposition approach, we derive exact autocorrelation and power-spectrum transmission laws that apply to arbitrarily complex upstream dynamics, including non-Markovian processes with memory effects. We further establish an oscillation-efficiency contraction principle and identify a necessary and sufficient condition for rhythm transmission, yielding an explicit critical degradation rate that separates rhythmic transmission from filtering. For a cyclic upstream driver, we derive an explicit relation revealing how molecular degradation, burst frequency, modulation depth, and upstream timing variability jointly govern rhythm transmission. Finally, we apply our theory to the p53 regulatory system, where the predicted degradation threshold is consistent with the experimentally observed distinction between pulsing and nonpulsing responses across ten target proteins.

\begin{figure}[htb!]
\centering\includegraphics[width=\linewidth]{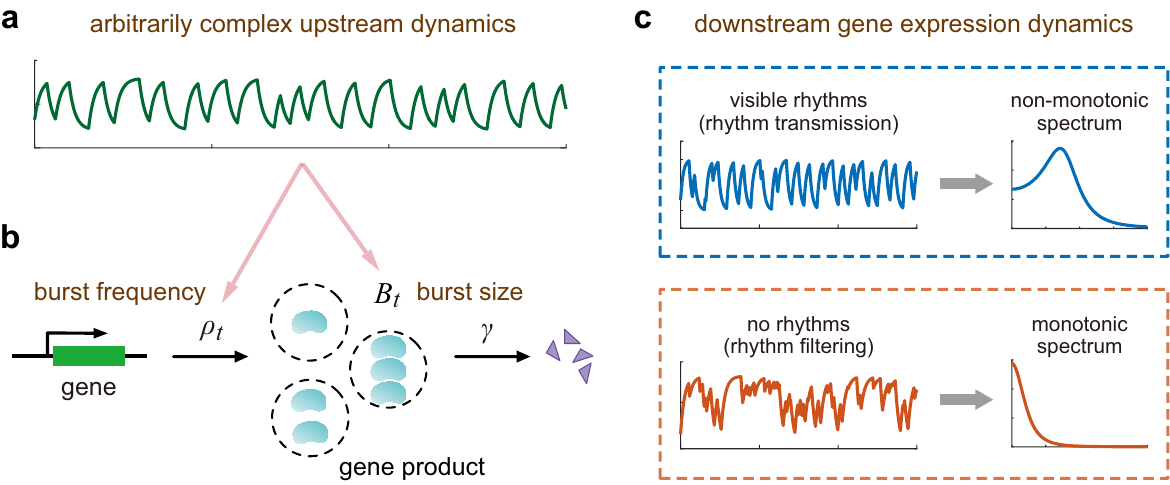}
\caption{\textbf{Stochastic processing and transmission of biological rhythms in bursty gene expression.}
\textbf{(a)} Representative trajectory of upstream rhythmic signals. The upstream dynamics can be arbitrarily complex, including stochasticity and memory effects.
\textbf{(b)} Bursty gene expression model. Both the burst frequency and the burst size can be modulated by the upstream signal.
\textbf{(c)} Representative trajectories and power spectra of downstream gene expression dynamics. Stochastic rhythms can be either transmitted or filtered.}
\label{figure1}
\end{figure}

\section{Results}

\subsection{Model}
Consider stochastic gene expression in a living cell driven by an upstream rhythmic signal. The upstream signal is modeled by an arbitrary stationary stochastic process $S_t$. The state space of $S_t$ can be either discrete or continuous. These states may represent regulator abundances \cite{hanson2019protein}, promoter configurations or activities \cite{zoller2015structure}, cell size \cite{thomas2021coordination}, or stages of a biological cycle \cite{jia2021frequency}. Here we do not require $S_t$ to be Markovian, allowing it to capture arbitrarily complex upstream dynamics, including stochasticity and memory effects (Fig.~\ref{figure1}(a)).

For downstream gene expression, we assume that the gene product (mRNA or protein) of interest is produced in bursts (Fig.~\ref{figure1}(b)). The burst sizes are assumed to follow an exponential distribution, consistent with experimental observations \cite{cai2006stochastic}. Both the burst frequency $\rho_t = \rho(S_t)$ and the mean burst size $B_t = B(S_t)$ may be modulated by the upstream rhythmic signal $S_t$. Conditional on the complete trajectory of the upstream signal, the cumulative gene product synthesis can be described by a stochastically modulated compound Poisson process
\begin{equation}
J_t = \sum_{n=1}^{N_t}\Delta_n,
\end{equation}
where $N_t$ is an inhomogeneous Poisson process with intensity $\rho_t$ that counts burst arrivals, and $\Delta_n$ denotes the size of the $n$th burst. Specifically, if a burst takes place at time $t$, then its size follows an exponential distribution with mean $B_t$. The burst sizes $\Delta_n$, $n\geq 1$ are conditionally independent given the complete upstream trajectory and the burst arrivals. Such compound Poisson representations have been widely used to model bursty stochastic gene expression \cite{friedman2006linking, pulkkinen2013distance, jkedrak2016time, bokes2017gene, jia2017emergent}.

The dynamics of gene product concentration is then governed by the stochastic differential equation
\begin{equation}\label{eq:model}
\mathrm dX_t=\mathrm dJ_t-\gamma X_{t-}\mathrm dt,
\end{equation}
where $J_t$ describes gene product synthesis and $\gamma>0$ is the degradation rate. In what follows, we assume that the system has reached the steady state and define
\begin{equation}\label{eq:uD}
U_t=\rho_t B_t,\qquad D=2\E[\rho_tB_t^2].
\end{equation}
Here $U_t$ is the conditional mean synthesis rate given the complete upstream trajectory and therefore represents the effective upstream drive experienced by the downstream gene product. Then Eq.~\eqref{eq:model} can be rewritten as
\begin{equation}\label{eq:martingale}
\mathrm dX_t=(U_t-\gamma X_{t-})\,\mathrm dt+\mathrm dM_t,
\end{equation}
where $M_t=J_t-\int_0^t U_s\dd s$ is a martingale \cite{anderson2011continuous}. Its predictable quadratic variation is $\langle M\rangle_t=2\int_0^t\rho_sB_s^2\dd s$, which implies that $\E\langle M\rangle_t=Dt$. Thus, $D$ quantifies the intrinsic variance injected into the gene expression dynamics per unit time, accounting for stochasticity in both burst arrivals and burst sizes.

We emphasize that here we focus on the concentration dynamics of the gene product rather than its copy number dynamics. The main reason is that, when the upstream signal represents the cell-cycle phase or cell size, the gene product copy number is approximately halved at cell division, resulting in a discontinuous reset. In contrast, such resetting can be neglected when considering gene product concentrations, since both the gene product copy number and cell volume are approximately halved at division, leaving the concentration essentially unchanged.

\subsection{Autocorrelation and power-spectrum transmission laws}
Stochastic rhythms are commonly described by autocorrelation functions and power spectra. For the upstream drive $U_t$ and the downstream gene product level $X_t$, their stationary autocorrelation functions are defined as $C_u(t)=\Cov(U_0,U_t)$ and $C(t)=\Cov(X_0,X_t)$, where $\Cov(\cdot,\cdot)$ denotes covariance. The corresponding power spectra are given by the Fourier transforms
\begin{equation*}
G_u(\omega)=\int_{\mathbb R}C_u(t)e^{-\ii\omega t}\dd t,
\quad G(\omega)=\int_{\mathbb R}C(t)e^{-\ii\omega t}\dd t,
\end{equation*}
where $\omega\geq 0$ denotes the angular frequency. A monotonically decreasing spectrum typically indicates the absence of a pronounced stochastic rhythm, whereas a nonmonotonic spectrum with a peak at a nonzero frequency provides a characteristic signature of stochastic oscillations (Fig.~\ref{figure1}(c)). The peak location identifies the dominant frequency. Throughout the following analysis, we assume that $\E[U_t^2]<\infty$, $D<\infty$, and $C_u(t)$ is absolutely integrable over $\mathbb R$. Under these conditions, both power spectra are bounded and continuous.

We next investigate how the power spectrum of the rhythmic upstream drive is transmitted to downstream gene expression. The martingale representation provides a direct route to deriving the transmission laws. Applying the method of variation of constants to Eq.~\eqref{eq:martingale}, we find that the downstream gene expression level can be decomposed as (see Appendix A)
\begin{equation}
X_t=Y_t+Z_t,
\end{equation}
where
\begin{equation}\label{eq:martingalerep}
Y_t=\int_{-\infty}^t e^{-\gamma(t-s)}U_s\dd s,\quad
Z_t=\int_{-\infty}^t e^{-\gamma(t-s)}\mathrm dM_s.
\end{equation}
Here $Y_t$ represents the exponentially filtered upstream drive, while $Z_t$ captures the filtered intrinsic fluctuations in gene expression. Conditional on the complete upstream trajectory, $Y_t$ is deterministic, and $Z_t$ has zero conditional mean by the martingale property of $M_t$. Hence,
\begin{equation}\label{eq:orthogonality}
\Cov(Y_0,Z_t)=\Cov(Z_0,Y_t)=0.
\end{equation}
Moreover, note that $Z_t=e^{-\gamma t}Z_0+\int_0^te^{-\gamma(t-s)}\mathrm dM_s$ for any $t\geq 0$. Using the martingale isometry, we obtain
\begin{equation}\label{eq:Zcov}
\Cov(Z_0,Z_t) = \Cov(Z_0,Z_{|t|})
=e^{-\gamma|t|}\E Z_0^2
=\frac{D}{2\gamma}e^{-\gamma|t|}.
\end{equation}
Since $X_t=Y_t+Z_t$, the orthogonality in Eq.~\eqref{eq:orthogonality} yields
\begin{equation}\label{eq:decomposition}
C(t)=\Cov(Y_0,Y_t)+\Cov(Z_0,Z_t).
\end{equation}
Thus, the upstream drive and intrinsic fluctuations contribute additively to the downstream autocorrelation function. It remains to evaluate the autocorrelation function of $Y_t$. Rewriting $Y_t$ as $Y_t=\int_0^\infty e^{-\gamma r}U_{t-r}\dd r$, and using the stationarity of $U_t$, we obtain
\begin{equation}\label{eq:Ycov}
\begin{split}
\Cov(Y_0,Y_t)
&=\int_0^\infty\!\!\int_0^\infty
e^{-\gamma(r+s)}C_u(t+r-s)\dd r\dd s\\
&=\frac{1}{2\gamma}\int_{\mathbb R}
e^{-\gamma|t-s|}C_u(s)\dd s.
\end{split}
\end{equation}
Combining Eqs.~\eqref{eq:Zcov}, \eqref{eq:decomposition}, and \eqref{eq:Ycov}, we obtain the following \emph{autocorrelation transmission law}:
\begin{equation}\label{eq:cov-transfer}
C(t)=\frac{1}{2\gamma}\int_{\mathbb R}
e^{-\gamma|t-s|}C_u(s)\dd s
+\frac{D}{2\gamma}e^{-\gamma|t|}.
\end{equation}
Taking the Fourier transform of both sides yields the corresponding \emph{power-spectrum transmission law}:
\begin{equation}\label{eq:psd-transfer}
G(\omega)=\frac{G_u(\omega)+D}
{\gamma^2+\omega^2}.
\end{equation}
Note that a related spectral transmission law has been derived in Ref.~\cite{song2019frequency} when the gene product is produced in a non-bursty manner; here we take bursting into account and allow both burst frequency and burst size to be modulated by the upstream drive. From Eq.~\eqref{eq:psd-transfer}, if the upstream spectrum $G_u(\omega)$ is monotonically decreasing for $\omega\geq 0$, then the downstream spectrum $G(\omega)$ is also monotonically decreasing. Similarly, if the upstream autocorrelation $C_u(t)$ is decreasing for $t\geq 0$, then so is the downstream autocorrelation $C(t)$ (see Appendix B).

\subsection{Transmission of the oscillation efficiency}
To further quantify the strength of stochastic rhythms, we define the oscillation efficiencies of the upstream and downstream signals as \cite{jia2024exact}
\begin{equation}\label{eq:eta}
\eta_u=1-\frac{G_u(0)}{G_u^{\max}},\qquad
\eta=1-\frac{G(0)}{G^{\max}},
\end{equation}
where $G_u^{\max}=\max_{\omega\geq0}G_u(\omega)$ and $G^{\max}=\max_{\omega\geq0}G(\omega)$ denote the spectral maxima. Clearly, both $\eta_u$ and $\eta$ lie between $0$ and $1$. A high oscillation efficiency indicates that the spectral power at the dominant frequency substantially exceeds the zero-frequency background, reflecting strong stochastic rhythms. In particular, if $G(\omega)$ is monotonically decreasing for $\omega\geq 0$, then $G^{\max}=G(0)$ and hence $\eta=0$. Thus, $\eta$ provides a quantitative measure of the strength of stochastic rhythms.

The transmission law in Eq.~\eqref{eq:psd-transfer} implies that
\begin{equation}
G^{\max}\leq\frac{G_u^{\max}+D}{\gamma^2},\qquad
G(0)=\frac{G_u(0)+D}{\gamma^2}.
\end{equation}
These relations yield the \emph{efficiency contraction principle}
\begin{equation}\label{eq:contraction}
\eta\leq
\frac{\eta_u}{1+\Rnoise(1-\eta_u)}
\leq\eta_u,
\end{equation}
where $\mathcal R=D/G_u(0)$ quantifies the strength of intrinsic gene expression fluctuations relative to the zero-frequency spectral power of the upstream rhythmic signal. We assume $G_u(0)>0$ whenever the readout-noise ratio $\mathcal R$ is introduced. In what follows, we refer to $\mathcal R$ as the \emph{readout noise}. The contraction principle shows that the oscillation efficiency of downstream gene expression can never exceed that of the upstream drive.

\subsection{Fast degradation induces rhythm transmission}
We next investigate how gene product degradation affects rhythm transmission. To this end, we define
\begin{equation}\label{eq:L}
L=\sup_{\omega>0}\frac{G_u(\omega)-G_u(0)}{\omega^2}\geq 0.
\end{equation}
Assume that the upstream signal has positive oscillation efficiency, or equivalently, $L>0$. For downstream gene expression to exhibit positive oscillation efficiency ($\eta>0$), there must exist some $\omega_*>0$ such that $G(\omega_*)>G(0)$. It then follows from Eq.~\eqref{eq:psd-transfer} that
\begin{equation}
\gamma^2\frac{G_u(\omega_*)-G_u(0)}{\omega_*^2}>G_u(0)+D.
\end{equation}
Taking the supremum over $\omega_*>0$ yields the following necessary and sufficient condition for rhythm transmission, valid for arbitrarily complex upstream processes:
\begin{equation}\label{eq:general-threshold}
\eta>0\quad\Longleftrightarrow\quad
\gamma^2L>G_u(0)+D.
\end{equation}
Since positive oscillation efficiency indicates the presence of pronounced stochastic rhythms, Eq.~\eqref{eq:general-threshold} establishes a critical degradation rate for rhythm transmission:
\begin{equation}\label{eq:dc-general}
\gamma_c
=\sqrt{\frac{G_u(0)+D}{L}}
=\sqrt{\frac{G_u(0)}{L}(1+\Rnoise)}.
\end{equation}
The downstream oscillation efficiency is positive if and only if $\gamma>\gamma_c$. Thus, rapid gene product degradation facilitates the transmission of stochastic rhythms from upstream signals to downstream gene expression. This provides a theoretical explanation for previous experimental observations that short-lived gene products track upstream rhythms, whereas long-lived products filter them out \cite{robertson2008real, cesbron2013light, porter2016p53, hanson2019protein, zambrano2016nf}.

\subsection{Efficiency bounds for Markovian drivers}
We now consider the special case where the upstream signal $S_t$ is modeled as an irreducible $N$-state Markov jump process with generator matrix $Q=(q_{ij})$ and stationary distribution $\pi=(\pi_i)$. Here $q_{ij}$ with $i\neq j$ denotes the transition rate from state $i$ to state $j$, and $q_{ii}=-\sum_{j\ne i}q_{ij}$. These states may represent promoter configurations \cite{zoller2015structure} or discrete stages of a biological cycle \cite{jia2021frequency}.

In particular, if the upstream dynamics satisfies detailed balance, i.e. $\pi_iq_{ij}=\pi_jq_{ji}$, then $C_u(t)$ is a nonnegative sum of decaying exponentials, and $G_u(\omega)$ is a nonnegative sum of Lorentzians \cite{qian2003fundamental}. In this case, both $C_u(t)$ and $G_u(\omega)$ are monotonically decreasing. Therefore, a microscopically reversible upstream signal cannot generate stochastic rhythms in downstream gene expression.

For an upstream signal with $N\geq 3$ states, a recent study obtained a universal upper bound on its oscillation efficiency, namely $\eta_u\leq\cos^2(\pi/N)$ \cite{jia2026fundamental}. Combining this bound with Eq.~\eqref{eq:contraction}, we obtain
\begin{equation}\label{eq:general-bound}
\eta\leq\frac{\cos^2(\pi/N)}{1+\Rnoise\sin^2(\pi/N)}.
\end{equation}
This establishes an upper bound on the oscillation efficiency of downstream gene expression, determined by the number of upstream states and the readout noise. For upstream systems with only one or two states, detailed balance necessarily holds, and hence $\eta=0$.

\begin{figure*}[htb!]
\centering\includegraphics[width=0.65\linewidth]{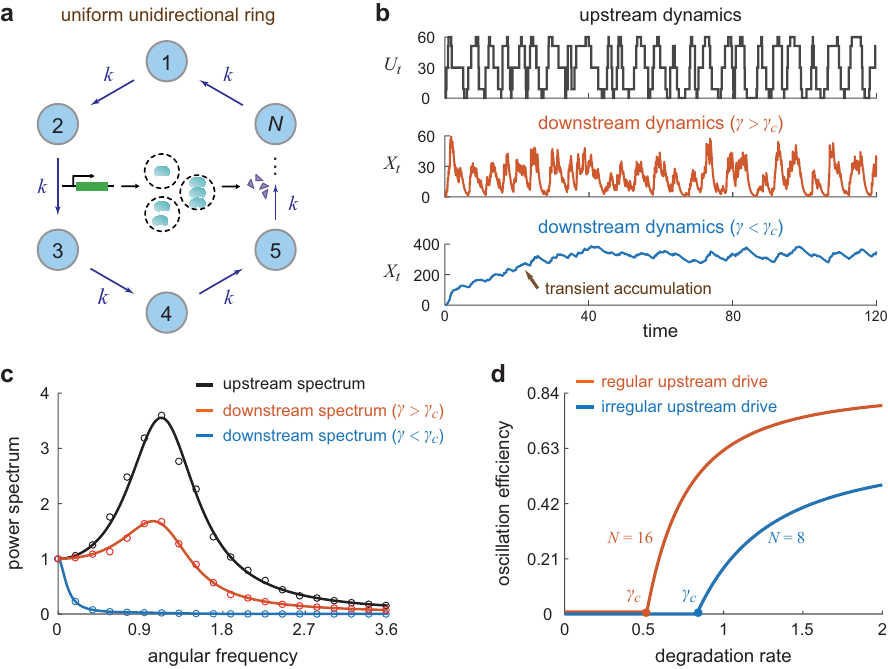}
\caption{\textbf{Transmission and filtering of stochastic rhythms in the frequency-modulated model.}
\textbf{(a)} The upstream rhythmic signal is modeled as an $N$-state uniform unidirectional ring.
\textbf{(b)} Representative trajectories of the upstream and downstream signals. Here the initial conditions are chosen to be $S_0=N/2$ and $X_0 = 0$.
\textbf{(c)} Power spectra of the upstream and downstream signals. Here the power spectra are normalized so that $G_u(0) = G(0) = 1$. Solid curves show the theoretical predictions obtained from Eqs.~\eqref{eq:psd-transfer} and \eqref{eq:ring-spec}, while circles represent simulation results estimated using Welch's method. Specifically, after discarding the initial transient and subtracting the stationary mean, each trajectory was divided into segments with 50\% overlap. A Hann window was applied to each segment, and the resulting periodograms were averaged.
\textbf{(d)} Oscillation efficiency $\eta$ as a function of the degradation rate $\gamma$.
In (b) and (c), the model parameters are chosen as $N=8$, $T_u=5$, $\bar\rho=10$, $\epsilon=1$, $B=3$, $\phi=0$, $\gamma=0.1$ for rhythm filtering, and $\gamma=1.5$ for rhythm transmission. In (d), we choose $N = 8,16$ and allow $\gamma$ to be varied; the other parameters are the same as in (b) and (c).}
\label{figure2}
\end{figure*}

\subsection{An explicit rhythm-transmission relation}
In Eq.~\eqref{eq:dc-general}, the critical degradation rate $\gamma_c$ depends on the upstream spectral quantities $G_u(0)$ and $L$, as well as the intrinsic noise strength $D$. However, it does not explicitly reveal how $\gamma_c$ depends on the underlying model parameters.

To make this dependence explicit, we specialize the upstream Markovian dynamics to an $N$-state uniform unidirectional ring with a common transition rate $k$, i.e. $q_{i,i+1}=k$ for all $i=1,\ldots,N$, where indices are understood modulo $N$ (Fig.~\ref{figure2}(a)). The period $\tau$ of the upstream rhythm, defined as the time required to complete one cycle around the ring, follows an Erlang distribution with mean $T_u:=\E\tau=N/k$ and squared coefficient of variation $\delta_u:=\Var(\tau)/(\E\tau)^2=1/N$, which quantifies the variability in cycle duration. Experimental studies have demonstrated that cell-cycle durations approximately follow Erlang distributions in various cell types \cite{chao2019evidence, perez2020effects}, while circadian measurements indicate relatively precise cycle timing \cite{phillips2021circadian}. These observations motivate the use of the uniform ring as an effective model of upstream rhythmic dynamics.

For analytical tractability, we further assume that the mean burst size $B$ is constant and independent of the upstream signal, while the burst frequency is harmonically modulated according to
\begin{equation}\label{eq:harmonic}
\rho(S_t)=\bar\rho\left[1+\epsilon\cos(2\pi S_t/N+\phi)\right],
\end{equation}
where $0<\epsilon\leq1$ denotes the modulation depth and $\phi$ denotes a phase offset. The assumption that burst size is independent of the upstream rhythmic signal is supported by experiments showing that burst frequency, rather than burst size, varies over the circadian cycle \cite{nicolas2018modulation, tong2025single} as well as the cell cycle \cite{padovan2015single, skinner2016single}. Measurements of p21 responses to p53 have also shown that activation probability is more sensitive to p53 levels than activation magnitude \cite{hafner2020quantifying}.

For this frequency-modulated model with $N\geq3$, direct calculations show that the upstream power spectrum is given by (see Appendix C)
\begin{equation}\label{eq:ring-spec}
G_u(\omega)=\frac{\bar\rho^2B^2\epsilon^2\mu(\omega^2+\mu^2+\nu^2)}{\omega^4+2(\mu^2-\nu^2)\omega^2+(\mu^2+\nu^2)^2},
\end{equation}
where $\mu=k[1-\cos(2\pi/N)]$ and $\nu=k\sin(2\pi/N)$. Hence,
\begin{equation}\label{eq:ring-R}
G_u(0)=\frac{\bar\rho^2B^2\epsilon^2}{2k},\qquad
D=2\bar\rho B^2.
\end{equation}
Inserting these expressions into Eq.~\eqref{eq:dc-general} yields an explicit formula for the critical degradation rate $\gamma_c$. For $N\leq3$, positive oscillation efficiency cannot be transmitted to downstream gene expression. For $N\geq4$, the critical degradation rate is given by (see Appendix D)
\begin{equation}\label{eq:ring-threshold}
\gamma_c = \sqrt{K_N(k)(1+\Rnoise)},
\end{equation}
where $\Rnoise=4k/(\bar\rho\epsilon^2)$ is the readout noise, and
\begin{equation*}
K_N(k)=\begin{cases}
2k^2,&N=4,\\
16k^2\sin^2\!\dfrac\pi N\cos\dfrac\pi N
\left(1-\cos\dfrac\pi N\right),&N\geq5.
\end{cases}
\end{equation*}
In the limit $N\gg1$, the variability in upstream cycle duration becomes small ($\delta_u\ll1$), and we obtain
\begin{equation}\label{eq:approx}
K_N(k)\approx\frac{8k^2\pi^4}{N^4}
=\frac{8\pi^4\delta_u^2}{T_u^2}.
\end{equation}
Combining Eqs.~\eqref{eq:ring-threshold} and \eqref{eq:approx}, we obtain the following explicit \emph{rhythm-transmission relation}:
\begin{equation}\label{eq:smallnoise}
\gamma>\gamma_c\approx
\frac{2\sqrt2\pi^2\delta_u}{T_u}\sqrt{1+\Rnoise},
\end{equation}
where $\Rnoise=4/(\bar\rho\epsilon^2T_u\delta_u)$ is the readout noise. This relation reveals how rhythm transmission to downstream gene expression depends on gene product degradation, gene expression bursting, and the period and variability of the upstream rhythm (Fig.~\ref{figure2}(b),(c)). In particular, it shows that rhythm transmission is facilitated by a larger degradation rate $\gamma$, a higher burst frequency $\bar\rho$, a larger modulation depth $\epsilon$, and lower variability $\delta_u$ in upstream cycle duration. Thus, a more regular upstream rhythm can be transmitted even when the downstream gene product degrades more slowly (Fig.~\ref{figure2}(d)).

Interestingly, Eq.~\eqref{eq:smallnoise} predicts that while increasing the burst frequency facilitates rhythm transmission, the mean burst size $B$ has no effect on it. To understand this, note that Eq.~\eqref{eq:psd-transfer} gives
\begin{equation}
\frac{G(\omega)}{G(0)}
=\frac{\gamma^2}{\gamma^2+\omega^2}
\frac{G_u(\omega)+D}{G_u(0)+D}.
\end{equation}
Eqs.~\eqref{eq:ring-spec} and \eqref{eq:ring-R} indicate that both $G_u(\omega)$ and $D$ scale as $B^2$. Consequently, the normalized downstream spectrum $G(\omega)/G(0)$ is independent of $B$. Note that this conclusion is not specific to the ring model. For arbitrarily complex upstream dynamics, uniformly rescaling the burst size $B_t$ by a factor $c>0$ scales both $G_u(\omega)$ and $D$ by $c^2$, leaving $G(\omega)/G(0)$ unchanged.

\subsection{Comparison with p53 experiments}
The p53 system provides a biological example in which a common pulsatile regulator elicits distinct downstream gene expression responses. Following DNA double-strand breaks, p53 accumulates in repeated pulses and activates downstream target genes involved in DNA repair, cell-cycle arrest, and apoptosis. An experimental study examined target protein responses to p53 pulsing in human MCF7 breast carcinoma cells and identified two broad response patterns \cite{hanson2019protein}: \emph{pulsing} proteins exhibit repeated rises and falls in response to p53 pulses, whereas \emph{nonpulsing} proteins accumulate with little detectable pulsation. Short-lived proteins generally preserve the upstream pulsatile rhythm, whereas longer-lived proteins tend to integrate successive pulses, thereby filtering out the rhythm (Fig.~\ref{figure2}(b)). Thus, targets of the same transcription factor can exhibit distinct temporal responses depending on their protein stability.

Moreover, Ref.~\cite{hanson2019protein} found that pulsing proteins typically have degradation rates above the p53 pulse frequency, whereas nonpulsing ones typically have degradation rates below it. We next use our rhythm-transmission relation to provide an explanation for this observation. In MCF7 cells, p53 pulses have a characteristic period of $T_u\approx5.5\,\mathrm{h}$ \cite{hanson2019protein}; in addition, earlier single-cell measurements of p53-Mdm2 dynamics also reported that the coefficient of variation of the pulse period is below $20\%$ in most oscillating cells \cite{geva2006oscillations}. We thus take $\delta_u=0.2^2=0.04$ as a representative estimate of the variability in cycle duration.

Proteins are typically abundant in mammalian cells, with reported median abundances of approximately $50{,}000$ protein molecules per cell in mouse fibroblasts and $18{,}000$ in HeLa cells \cite{schwanhausser2011global, nagaraj2011deep}. Under large-system scaling, intrinsic gene expression fluctuations decrease with increasing protein abundance \cite{jia2019single}. These high protein abundances motivate the small-readout-noise approximation $\Rnoise\ll1$, provided that burst frequency is sufficiently high and the modulation depth is appreciable (see Supplemental Material \cite{supp} for details). Under this approximation, the critical degradation rate for rhythm transmission becomes
\begin{equation}
\gamma_c\approx
\frac{2\sqrt2\pi^2\delta_u}{T_u}
\approx\frac{1.12}{T_u}
=0.203\;\mathrm{h}^{-1}.
\end{equation}
Remarkably, this threshold is close to the upstream p53 pulse frequency $1/T_u\approx0.182\;\mathrm{h}^{-1}$. Table~\ref{tab:protein} compares the predicted threshold with the degradation rates of all ten proteins measured in Ref.~\cite{hanson2019protein}. Strikingly, all four pulsing proteins have degradation rates above the predicted threshold, whereas all six nonpulsing proteins have degradation rates below it.

\begin{table}[htb!]
\caption{Comparison of the theoretical critical degradation rate $\gamma_c\approx 0.203\;\mathrm{h}^{-1}$ with the degradation rates of ten p53 target proteins, which are taken from Table S4 of Ref.~\cite{hanson2019protein}. The observed pulsing and nonpulsing classifications are based on the authors' population-level measurements. The degradation rate of TIGAR was set to zero in Ref.~\cite{hanson2019protein} because no measurable decay was observed over the experimental time window. In addition, the initial induction of EGFR precedes p53 activation in the experiments, suggesting that its response may involve a p53-independent component. Thus, EGFR provides a less direct test of the proposed rhythm-transmission mechanism.}\label{tab:protein}
\begin{ruledtabular}\begin{tabular}{llll}
Protein & $\gamma\;(\mathrm h^{-1})$ & Position & Observed\\
\hline
GDF15 &1.177 &Above $\gamma_c$ &Pulsing\\
MDM2  &0.792 &Above $\gamma_c$ &Pulsing\\
p21   &0.255 &Above $\gamma_c$ &Pulsing\\
WIP1  &0.226 &Above $\gamma_c$ &Pulsing\\
EGFR  &0.088 &Below $\gamma_c$ &Nonpulsing\\
PUMA  &0.056 &Below $\gamma_c$ &Nonpulsing\\
BAX   &0.026 &Below $\gamma_c$ &Nonpulsing\\
PIG3  &0.006 &Below $\gamma_c$ &Nonpulsing\\
AMPKB1&0.003 &Below $\gamma_c$ &Nonpulsing\\
TIGAR &0.000 &Below $\gamma_c$ &Nonpulsing
\end{tabular}\end{ruledtabular}
\end{table}

\section{Conclusions and discussion}
In this study, we have developed a general analytical framework for stochastic rhythm transmission in bursty gene expression based on a martingale-based orthogonal decomposition approach. By separating fluctuations transmitted from the upstream regulatory signal from intrinsic noise generated by stochastic bursting, we derived exact autocorrelation and power-spectrum transmission laws for arbitrarily complex upstream dynamics, including non-Markovian processes with memory effects. These results further establish an oscillation-efficiency contraction principle and a necessary and sufficient condition for rhythm transmission, thereby providing a quantitative framework for understanding how stochastic rhythms are transmitted or filtered during gene expression. Importantly, the framework accommodates upstream regulation of both burst frequency and burst size, extending the applicability of spectral approaches to a broad class of stochastic gene expression models.

Our results reveal several fundamental principles governing stochastic rhythm transmission. First, downstream gene expression cannot exhibit a higher oscillation efficiency than its upstream drive, indicating that stochastic bursting and molecular degradation impose intrinsic limitations on the fidelity of rhythm transmission. Second, sufficiently rapid degradation is required for downstream gene products to preserve upstream rhythms, whereas slow degradation leads to temporal integration and rhythm filtering. Third, rhythm transmission depends not only on molecular degradation but also on the statistical properties of gene expression bursting and upstream timing. In particular, increasing burst frequency or modulation depth facilitates rhythm transmission, while reducing variability in upstream cycle duration allows rhythms to be transmitted even by more slowly degrading gene products. Interestingly, uniformly rescaling burst sizes at fixed burst frequency leaves the normalized downstream power spectrum and oscillation efficiency unchanged. This distinction between burst frequency and burst size highlights the importance of how gene expression is regulated, rather than simply how much gene product is synthesized.

The framework also provides a quantitative connection between molecular turnover and experimentally observed gene expression dynamics. In the p53 regulatory system, different target proteins exhibit markedly different temporal responses despite being driven by the same pulsatile regulator. Our predicted critical degradation rate is consistent with the reported pulsing and nonpulsing classifications of all ten target proteins in Ref.~\cite{hanson2019protein}, suggesting that differences in protein stability contribute to their distinct temporal responses. This agreement is particularly notable because the threshold is estimated from the characteristic period and timing variability of upstream p53 oscillations rather than fitted to the degradation rates of individual target proteins. Nevertheless, the experimental classifications are based on transient population-level responses, whereas our oscillation efficiency is defined through stationary power spectra. Thus, the comparison provides supporting evidence for the proposed transmission criterion, while direct single-cell measurements of upstream and downstream power spectra would offer a more stringent quantitative test.

Several extensions of the present framework merit further investigation. Our analysis assumes stationary upstream signals, conditionally Poisson burst arrivals, and first-order degradation of gene products. Although these assumptions provide a useful description of many gene expression systems, cellular regulation may involve non-stationary signals, history-dependent synthesis, nonlinear degradation, or feedback from downstream gene products to upstream regulators. Extending the present approach to such settings would help determine which transmission principles remain valid beyond linear downstream dynamics. Another important direction is to investigate how stochastic rhythms propagate through multistage regulatory cascades, such as transcription, translation, and post-translational regulation, where different molecular lifetimes and noise sources may jointly shape the final response.

More broadly, our findings suggest potential strategies for controlling rhythmic gene expression in natural and synthetic regulatory systems. By tuning molecular degradation, burst frequency, or the regularity of upstream oscillations, it may be possible to enhance the transmission of biologically relevant rhythms or suppress unwanted fluctuations. The analytical framework developed here provides a starting point for investigating such strategies and for connecting molecular kinetic parameters to the temporal organization of gene expression. Understanding these connections may help clarify how cells selectively preserve, attenuate, or integrate rhythmic regulatory information across different molecular timescales.

\section{Acknowledgments}
The author acknowledges support from the National Natural Science Foundation of China with Grant Nos. 12526215, T2541031, and 12271020.

\section{Data and code availability}
The MATLAB code used to generate Fig.~\ref{figure2} is available at
\url{https://github.com/chenjiahit/stochastic-rhythm-transmission}. The experimental data used in Table~\ref{tab:protein} were obtained from Ref.~\cite{hanson2019protein}.

\section{Appendices}

\subsection{Appendix A: Decomposition of downstream gene expression dynamics}\label{appendixA}
Here we compute the explicit stationary solution of Eq.~\eqref{eq:martingale} using stochastic analysis techniques and the method of variation of constants. We assume that the upstream driving is stationary and satisfies
\begin{equation}
\E U_t^2<\infty,
\qquad
D=2\E[\rho_tB_t^2]<\infty.
\end{equation}
For the stationary representation, the upstream process and its conditionally Poisson burst process are understood to be defined
on the whole real line. Stochastic integrals over finite intervals are defined using the corresponding compensated increments.

Under the model assumptions, $M_t$ is a square-integrable martingale on every finite interval, relative to the filtration containing the complete upstream trajectory and all burst events and marks observed up to the current time. Its predictable quadratic
variation satisfies
\begin{equation}\label{eq:app-bracket}
\mathrm d\langle M\rangle_t
=2\rho_tB_t^2\dd t,
\qquad
\E[\mathrm d\langle M\rangle_t]=D\dd t.
\end{equation}

Fix a finite initial time $a<t$ and apply It\^o's formula
for semimartingales with jumps to
\begin{equation}
F(t,x)=e^{\gamma t}x.
\end{equation}
Since $F$ is linear in $x$, its second derivative with respect to $x$ vanishes. Moreover, the jump correction vanishes because
\begin{equation}
F(s,X_s)-F(s,X_{s-})
-\partial_xF(s,X_{s-})\Delta X_s=0.
\end{equation}
Consequently,
\begin{equation}
\begin{split}
\mathrm d(e^{\gamma t}X_t)
&=\gamma e^{\gamma t}X_{t-}\dd t
  +e^{\gamma t}\mathrm dX_t \notag\\
&=e^{\gamma t}U_t\dd t
  +e^{\gamma t}\mathrm dM_t.
\end{split}
\end{equation}
Integrating from $a$ to $t$ and multiplying by $e^{-\gamma t}$ gives
\begin{equation}\label{eq:app-finite-time}
\begin{split}
X_t=&\;e^{-\gamma(t-a)}X_a
+\int_a^t e^{-\gamma(t-s)}U_s\dd s\\
&\;+\int_a^t e^{-\gamma(t-s)}\mathrm dM_s.
\end{split}
\end{equation}

We next justify taking $a\to-\infty$. On the one hand, deterministic convolution
\begin{equation}
Y_t=\int_{-\infty}^t e^{-\gamma(t-s)}U_s\dd s
\end{equation}
is defined as an $L^2$ limit of finite-interval integrals. Indeed, the weighted Cauchy-Schwarz inequality and stationarity imply
\begin{equation}
\begin{split}
&\E\left|
\int_{-\infty}^a e^{-\gamma(t-s)}U_s\dd s
\right|^2\\
&\leq
\left(\int_{-\infty}^a e^{-\gamma(t-s)}\dd s\right)
\left(\int_{-\infty}^a
e^{-\gamma(t-s)}\E U_s^2\dd s\right)
\notag\\
&=
\frac{\E U_0^2}{\gamma^2}
e^{-2\gamma(t-a)}
\longrightarrow0,\qquad
a\to-\infty.
\end{split}
\end{equation}
On the other hand, the stochastic convolution
\begin{equation}
Z_t=\int_{-\infty}^t e^{-\gamma(t-s)}\mathrm dM_s
\end{equation}
is defined similarly as an $L^2$ limit of finite-interval integrals. Indeed, for $b<a<t$, the martingale isometry gives
\begin{equation}
\begin{split}
\E\left|
\int_b^a e^{-\gamma(t-s)}\mathrm dM_s
\right|^2
&=\E\int_b^a e^{-2\gamma(t-s)}
  \mathrm d\langle M\rangle_s\\
&=D\int_b^a e^{-2\gamma(t-s)}\dd s\\
&\leq \frac{D}{2\gamma}e^{-2\gamma(t-a)}.
\end{split}
\end{equation}
The finite-interval integrals therefore form an $L^2$-Cauchy family as $b\to-\infty$. In particular,
\begin{equation}\label{eq:app-Z-moments}
\E Z_t=0,
\qquad
\E Z_t^2=\frac{D}{2\gamma}.
\end{equation}
Finally, for a stationary solution with finite second moment, the initial term in Eq.~\eqref{eq:app-finite-time} satisfies
\begin{equation}
\E\left|e^{-\gamma(t-a)}X_a\right|^2
=e^{-2\gamma(t-a)}\E X_0^2
\longrightarrow0.
\end{equation}
Taking the $L^2$ limit in Eq.~\eqref{eq:app-finite-time} therefore yields the stationary solution $X_t = Y_t+Z_t$.
Conversely, these convergent convolutions construct the stationary finite-second-moment solution.

%\paragraph{Properties of the two components.}
%The upstream component satisfies
%\begin{equation}
%Y_t-Y_a=\int_a^t(U_s-\gamma Y_s)\dd s.
%\end{equation}
%Hence $Y$ is locally absolutely continuous and has finite
%variation on every compact time interval.
%
%The intrinsic component satisfies
%\begin{equation}\label{eq:app-Z-dynamics}
%\mathrm dZ_t=-\gamma Z_t\dd t+\mathrm dM_t.
%\end{equation}
%In particular, for $a<t$,
%\begin{equation}
%Z_t=e^{-\gamma(t-a)}Z_a
%+\int_a^t e^{-\gamma(t-s)}\mathrm dM_s,
%\end{equation}
%and therefore
%\begin{equation}
%\E[Z_t\mid\mathcal F_a]
%=e^{-\gamma(t-a)}Z_a.
%\end{equation}
%Thus $Z$ is generally not a martingale: it is a stationary,
%exponentially filtered martingale fluctuation.
%
%Finally, let
%$\mathcal S=\sigma(S_s:s\in\mathbb R)$
%denote the complete upstream trajectory.
%Conditional on $\mathcal S$, the compensated burst integrals
%have zero mean. Passing to the $L^2$ limit gives
%\begin{equation}
%\E[Z_t\mid\mathcal S]=0.
%\end{equation}
%Since $Y_s$ is $\mathcal S$-measurable for every $s$, it follows that
%\begin{equation}
%\Cov(Y_s,Z_t)
%=\E\!\left[
%(Y_s-\E Y_s)\E[Z_t\mid\mathcal S]
%\right]=0
%\end{equation}
%for all $s,t\in\mathbb R$.
%This establishes the orthogonality used in the covariance
%transmission law.

\subsection{Appendix B: Monotonicity of the autocorrelation function}\label{appendixB}
In the main text, we have demonstrated that if the upstream autocorrelation function $C_u(t)$ is monotonically decreasing for $t\geq0$, then the downstream autocorrelation function $C(t)$ is also decreasing on this interval.

Since $U_t = \rho_tB_t$ is a real-valued stationary process, its autocorrelation function is even:
\begin{equation}\label{eq:Cu-even}
C_u(-t)=C_u(t).
\end{equation}
Moreover, the Cauchy-Schwarz inequality gives
\begin{equation}
|C_u(t)|\leq C_u(0)<\infty.
\end{equation}
The autocorrelation function transmission law reads
\begin{equation}\label{eq:C-transfer-appB}
C(t)=F(t)+\frac{D}{2\gamma}e^{-\gamma|t|},
\end{equation}
where
\begin{equation}
F(t)=\frac{1}{2\gamma}
\int_{\mathbb R}e^{-\gamma|t-s|}C_u(s)\,\dd s.
\end{equation}
Since $e^{-\gamma|t|}$ is monotonically decreasing for $t\geq 0$, we only need to prove that $F(t)$ is decreasing for $t\geq0$. In fact, differentiating the exponential kernel gives
\begin{equation}\label{eq:F-derivative-appB}
\begin{aligned}
F'(t)
={}&-\frac12\int_{-\infty}^{t}
e^{-\gamma(t-s)}C_u(s)\,\dd s\\
&+\frac12\int_{t}^{\infty}
e^{-\gamma(s-t)}C_u(s)\,\dd s\\
={}&\frac12\int_0^\infty e^{-\gamma r}
\bigl[C_u(t+r)-C_u(t-r)\bigr]\,\dd r.
\end{aligned}
\end{equation}
Differentiation under the integral is justified by the boundedness of $C_u(t)$ and the integrability of the
derivative of the exponential kernel. For any $t,r\geq0$, we have
\begin{equation}
t+r\geq |t-r|.
\end{equation}
Combining the assumed monotonicity of $C_u(t)$ for $t\geq 0$ with its evenness therefore yields
\begin{equation}
C_u(t+r)\leq C_u(|t-r|)=C_u(t-r).
\end{equation}
Consequently, the integrand in Eq.~\eqref{eq:F-derivative-appB} is nonpositive, implying $F'(t)\leq0$ for $t\geq0$. This gives the desired result.

\subsection{Appendix C: Derivation of the upstream power spectrum for the frequency-modulated model}\label{appendixC}
We consider an upstream process $S_t$ on a unidirectional ring with $N$ states, labeled $i=0,\ldots,N-1$. Transitions from $i$ to $i+1$ modulo $N$ occur at rate $k$, and the stationary distribution is uniform. The burst frequency is
\begin{equation}
\rho(i)=\bar\rho
\left[1+\epsilon\cos(\theta i+\phi)\right],
\qquad
\theta=\frac{2\pi}{N},
\end{equation}
where $0<\epsilon\leq1$. The burst sizes are exponentially distributed with a state-independent mean $B$. Thus, the upstream driving is $U_t=B\rho(S_t)$.

We first derive the upstream spectrum for $N\geq3$, and then calculate the critical degradation rate. Let
\begin{equation}
A=\bar\rho^2B^2\epsilon^2,
\qquad
\mu=k(1-\cos\theta),
\qquad
\nu=k\sin\theta.
\end{equation}
For $f(i)=e^{\mathrm i\theta i}$, the generator $Q$
of the ring satisfies
\begin{equation}
(Qf)(i)
=k[f(i+1)-f(i)]
=(-\mu+\mathrm i\nu)f(i).
\end{equation}
Consequently, for $t\geq0$,
\begin{equation}
\mathbb E\!\left[
e^{\mathrm i\theta S_t}\mid S_0=i
\right]
=e^{(-\mu+\mathrm i\nu)t}
e^{\mathrm i\theta i}.
\end{equation}
Taking real parts yields
\begin{equation}
\mathbb E\!\left[
\cos(\theta S_t+\phi)\mid S_0=i
\right]
=e^{-\mu t}\cos(\theta i+\phi+\nu t).
\end{equation}

For $N\geq3$, averaging over the uniform stationary distribution gives
\begin{equation}
\frac1N\sum_{i=0}^{N-1}
\cos(\theta i+\phi)=0
\end{equation}
and
\begin{equation}
\frac1N\sum_{i=0}^{N-1}
\cos(\theta i+\phi)
\cos(\theta i+\phi+\nu t)
=\frac12\cos(\nu t).
\end{equation}
It follows that $\mathbb E U_t=\bar\rho B$ and
\begin{equation}\label{eq:ring-cov-appC}
C_u(t)=\frac{A}{2}
e^{-\mu|t|}\cos(\nu t).
\end{equation}
Using the Fourier-transform convention
\begin{equation}
G_u(\omega)=
\int_{\mathbb R}C_u(t)e^{-\mathrm i\omega t}\,\dd t,
\end{equation}
we obtain
\begin{equation}
G_u(\omega)
=\frac{A\mu}{2}
\left[
\frac{1}{\mu^2+(\omega-\nu)^2}
+
\frac{1}{\mu^2+(\omega+\nu)^2}
\right].
\end{equation}
Combining the two fractions gives
\begin{equation}\label{eq:ring-spectrum-appC}
G_u(\omega)
=
\frac{A\mu(\omega^2+\mu^2+\nu^2)}
{\omega^4+2(\mu^2-\nu^2)\omega^2
+(\mu^2+\nu^2)^2}.
\end{equation}

\subsection{Appendix D: Derivation of the explicit rhythm-transmission relation}\label{appendixD}
Since $\mu^2+\nu^2=2k\mu$, we have
\begin{equation}
G_u(0)
=\frac{A\mu}{\mu^2+\nu^2}
=\frac{A}{2k}.
\end{equation}
Moreover, the second moment of an exponential burst size is $2B^2$, so $D=2\mathbb E[\rho(S_t)]B^2=2\bar\rho B^2$. Therefore, the readout noise is given by
\begin{equation}\label{eq:ring-noise-appC}
\Rnoise = \frac{D}{G_u(0)}
=\frac{4k}{\bar\rho\epsilon^2}.
\end{equation}
In the main text, we have also proved that the critical degradation rate for rhythm transmission is given by
\begin{equation}\label{eq:ring-critical-appC}
\gamma_c=\sqrt{\frac{G_u(0)}{L}(1+\Rnoise)}.
\end{equation}
Positive efficiency occurs if and only if $\gamma>\gamma_c$. To calculate $L$, we introduce
\begin{equation*}
z=\omega^2,\quad
c=\mu^2+\nu^2,\quad
b=\mu^2-\nu^2,\quad
a=3\nu^2-\mu^2.
\end{equation*}
Eq.~\eqref{eq:ring-spectrum-appC} then gives
\begin{equation}
\frac{G_u(\omega)}{G_u(0)}
=\frac{c(z+c)}{z^2+2bz+c^2}.
\end{equation}
Subtracting one and dividing by $z$ yields
\begin{equation}
\frac{G_u(\omega)-G_u(0)}
{\omega^2G_u(0)}
=f(z),
\quad
f(z)=\frac{a-z}{z^2+2bz+c^2}.
\end{equation}
Hence $L=G_u(0)\sup_{z>0}f(z)$. The denominator is strictly positive because
\begin{equation}
z^2+2bz+c^2
=(z+\mu^2-\nu^2)^2+4\mu^2\nu^2.
\end{equation}

For $N\leq 2$, the upstream driving is reversible and hence the downstream power spectrum cannot have positive oscillation efficiency. For $N=3$, we have
\begin{equation}
\mu=\frac{3k}{2},
\qquad
\nu=\frac{\sqrt3\,k}{2},
\qquad
a=3\nu^2-\mu^2=0.
\end{equation}
Therefore, $f(z)<0$ for all $z>0$, so $G_u(\omega)<G_u(0)$ for every $\omega>0$. Therefore, positive oscillation efficiency cannot
be transmitted for $N\leq3$.

For $N\geq4$, we have $a>0$. Since $f(z)>0$ precisely when $0\leq z<a$, its positive maximum can be found on this interval.
Direct differentiation gives
\begin{equation}\label{eq:fprime-appC}
f'(z)=
\frac{(z-a)^2-4\nu^2c}
{(z^2+2bz+c^2)^2}.
\end{equation}
On $0\leq z<a$, the numerator changes sign at most once, at
\begin{equation}
z_*=a-2\nu\sqrt c.
\end{equation}
If $z_*\leq0$, the supremum is approached as $z\downarrow0$, giving
\begin{equation}\label{eq:fmax-boundary-appC}
\sup_{z>0}f(z)=f(0)=\frac{a}{c^2}.
\end{equation}
If $z_*>0$, then $f$ increases up to $z_*$ and decreases thereafter within its positive region. In this case,
\begin{equation}\label{eq:fmax-interior-appC}
\sup_{z>0}f(z)
=f(z_*)
=\frac{1}{4\nu(\sqrt c-\nu)}.
\end{equation}
The last identity follows from the stationarity
condition
\begin{equation}
z_*^2+2bz_*+c^2
=-(a-z_*)(2z_*+2b)
\end{equation}
and
\begin{equation}
2z_*+2b=-4\nu(\sqrt c-\nu).
\end{equation}
To determine which case applies, write
\begin{equation}
s=\sin\frac{\pi}{N},
\qquad
q=\cos\frac{\pi}{N}.
\end{equation}
Then
\begin{equation}
\mu=2ks^2,\qquad
\nu=2ksq,\qquad
\sqrt c=2ks,
\end{equation}
and
\begin{equation}
z_*=4k^2s^2(4q^2-2q-1).
\end{equation}
Since
\begin{equation}
4q^2-2q-1=0
\quad\Longleftrightarrow\quad
q=\frac{1+\sqrt5}{4}
=\cos\frac{\pi}{5},
\end{equation}
we have $z_*<0$ for $N=4$, $z_*=0$ for $N=5$, and $z_*>0$ for $N\geq6$.

For $N=4$, we have $\mu=\nu=k$, so $a=2k^2$ and $c=2k^2$. Eq.~\eqref{eq:fmax-boundary-appC} gives
\begin{equation}
\left[\sup_{z>0}f(z)\right]^{-1}=2k^2.
\end{equation}
For $N=5$, we have $z_*=0$, and the expressions in Eqs.~\eqref{eq:fmax-boundary-appC} and \eqref{eq:fmax-interior-appC} coincide. For $N\geq6$, Eq.~\eqref{eq:fmax-interior-appC} gives
\begin{equation}
\left[\sup_{z>0}f(z)\right]^{-1}
=4\nu(\sqrt c-\nu)
=16k^2s^2q(1-q).
\end{equation}
Therefore, defining
\begin{equation}
K_N(k)=
\left[\sup_{z>0}f(z)\right]^{-1},
\end{equation}
we obtain
\begin{equation*}
K_N(k)=
\begin{cases}
2k^2,&N=4,\\[2mm]
16k^2\sin^2\!\dfrac{\pi}{N}
\cos\dfrac{\pi}{N}
\left(1-\cos\dfrac{\pi}{N}\right),
&N\geq5.
\end{cases}
\end{equation*}
Substituting $L=G_u(0)/K_N(k)$ and Eq.~\eqref{eq:ring-noise-appC} into Eq.~\eqref{eq:ring-critical-appC} finally yields
\begin{equation}
\gamma_c=
\sqrt{K_N(k)
\left(1+\frac{4k}{\bar\rho\epsilon^2}\right)}.
\end{equation}
Thus, for $N\geq4$, the downstream process has positive oscillation efficiency if and only if $\gamma>\gamma_c$. At $\gamma=\gamma_c$, its oscillation efficiency is zero.

% Intentionally left blank at the author's request.
\footnotesize\setlength{\bibsep}{5pt}
\bibliography{references}
\bibliographystyle{apsrev4-2}

\end{document}

%\linenumbers

%%%%%%%%%% Theorem Enviroment %%%%%%%%%%
\makeatletter
\thm@headfont{\bf}
\makeatother

\theoremstyle{definition}
\newtheorem{theorem}{Theorem}[section]
\newtheorem{definition}[theorem]{Definition}
\newtheorem{lemma}[theorem]{Lemma}
\newtheorem{proposition}[theorem]{Proposition}
\newtheorem{corollary}[theorem]{Corollary}
\newtheorem{remark}[theorem]{Remark}
\newtheorem{example}[theorem]{Example}

%%%%%%%%%% Page styles %%%%%%%%%%
\lhead{}
\rhead{}
\lfoot{}
\rfoot{}

%%%%%%%%%% Abbreviations %%%%%%%%%%
\renewcommand{\refname}{References}
\renewcommand{\figurename}{Figure}
\renewcommand{\tablename}{Table}
\renewcommand{\proofname}{Proof}

%%%%%%%%%% Special symbols %%%%%%%%%%
\newcommand{\tr}{\mathrm{tr}}
\newcommand{\re}{\mathrm{Re}}
\newcommand{\one}{\mathbbm{1}}
\newcommand{\Pnum}{\mathbb{P}}
\newcommand{\Enum}{\mathbb{E}}
\newcommand{\Rnum}{\mathbb{R}}
\newcommand{\dnum}{\mathrm{d}}
\newcommand{\hyper}{{}_2F_1}
\newcommand{\confl}{{}_1F_1}
\newcommand{\ip}[2]{\langle #1,#2\rangle_\pi}
\newcommand{\HH}{\mathcal H_0}
\newcommand{\EE}{\mathcal E}
\newcommand{\Var}{\operatorname{Var}}
\newcommand{\Rea}{\operatorname{Re}}
\newcommand{\artanh}{\operatorname{artanh}}
\newcommand{\diag}{\operatorname{diag}}
\newcommand{\dd}{\,\mathrm d}
\newcommand{\ii}{\mathrm i}
\newcommand{\Rnoise}{\mathcal R}

%%%%%%%%%% Title and authors %%%%%%%%%%
\title{\Large{\textbf{Supplemental Material} ---\\
Transmission and filtering of biological rhythms in stochastic gene expression}}
\author{Chen Jia}

\date{}
\maketitle
\tableofcontents
\thispagestyle{empty}

%%%%%%%%%% Page Styles %%%%%%%%%%%
\paperfont

\captionsetup{font=footnotesize}

%%%%%%%%%% Abstract %%%%%%%%%%
%\begin{abstract}

%\end{abstract}
\newpage

%%%%%%%%%% Main Text %%%%%%%%%%

\section{Estimation of the protein readout noise in mammalian cells}
In the main text, we have shown that the readout noise in the frequency-modulated model is given by
\begin{equation}\label{eq:R-appD}
\Rnoise=\frac{D}{G_u(0)}
=\frac{4}{\bar\rho\epsilon^2T_u\delta_u}.
\end{equation}
We first explain its scaling in the large-system limit and then estimate its magnitude using reported mammalian protein abundances.

\subsection{Large-system scaling}
Let $\Omega$ denote the system-size parameter, and let $X_t^{(\Omega)}$ be the protein abundance normalized by $\Omega$. We assume that the burst frequency scales as
\begin{equation}
\rho_t^{(\Omega)}=\Omega r_t,
\end{equation}
whereas the burst size in molecular units has a fixed exponential distribution with mean $b$. The corresponding mean jump size of the normalized abundance is therefore $b/\Omega$. Under this scaling, the mean synthesis rate is
\begin{equation}
U_t^{(\Omega)}
=\rho_t^{(\Omega)}\frac{b}{\Omega}
=r_tb,
\end{equation}
which is independent of $\Omega$. In contrast, the intrinsic noise intensity is
\begin{equation}
D^{(\Omega)}
=2\mathbb E[\rho_t^{(\Omega)}]
\left(\frac{b}{\Omega}\right)^2
=\frac{2\bar r b^2}{\Omega}.
\end{equation}
Thus, for fixed upstream dynamics and modulation depth, the upstream spectrum remains unchanged, whereas
\begin{equation}\label{eq:R-scaling-appD}
\Rnoise^{(\Omega)}
=\frac{4}{\Omega\bar r\epsilon^2T_u\delta_u}
=O(\Omega^{-1}).
\end{equation}

The normalized intrinsic fluctuations consequently vanish in the large-system limit, and the downstream dynamics approach
\begin{equation}
\frac{\dd x_t}{\dd t}=r_tb-\gamma x_t.
\end{equation}
This equation is deterministic conditional on the upstream trajectory; stochastic upstream fluctuations are retained.

\subsection{Expression in terms of protein abundance}
To connect Eq.~\eqref{eq:R-appD} to abundance data, let $\bar n$ denote the mean protein abundance in molecular units, corresponding to the mean concentration in the effective stationary model. Stationary production--loss balance gives
\begin{equation}
\bar\rho b=\gamma\bar n.
\end{equation}
Therefore,
\begin{equation}\label{eq:R-abundance-appD}
\Rnoise
=\frac{4b}
{\gamma\bar n\epsilon^2T_u\delta_u}.
\end{equation}
Here, $\gamma\bar n$ is the mean production flux, whereas $\bar\rho=\gamma\bar n/b$ is the burst frequency.

Eq.~\eqref{eq:R-abundance-appD} makes the large-system scaling explicit: at fixed $b$, $\gamma$, and upstream parameters, increasing protein abundance reduces the readout noise. The relevant abundance scale is $\bar n/b$, rather than $\bar n$ alone.

\subsection{Estimates based on mammalian protein abundances}
The corrected proteomic measurements of Schwanh\"ausser \emph{et al.} reported a median protein abundance of approximately $5.0\times10^4$ molecules per cell in mouse NIH3T3 fibroblasts \cite{schwanhausser2011global}. Nagaraj \emph{et al.} reported a median abundance of approximately $1.8\times10^4$ molecules per cell in HeLa cells \cite{nagaraj2011deep}. We use these reported proteome-wide median abundances as representative proxies for the mean abundance $\bar n$ in Eq.~
\eqref{eq:R-abundance-appD}, rather than as measurements of individual p53 target proteins.

Taking $T_u=5.5\,\mathrm h$ and $\delta_u=0.04$, Eq.~\eqref{eq:R-abundance-appD} becomes
\begin{equation}
\Rnoise
=\frac{4b}{0.22\,\mathrm h\,
\gamma\bar n\epsilon^2}.
\end{equation}
For mouse fibroblast abundance levels, this gives
\begin{equation}\label{eq:R-mouse-appD}
\Rnoise
\simeq1.82\times10^{-3}\,b
\left(\frac{0.2\,\mathrm h^{-1}}{\gamma}\right)
\epsilon^{-2},
\end{equation}
whereas for HeLa abundance levels,
\begin{equation}\label{eq:R-HeLa-appD}
\Rnoise
\simeq5.05\times10^{-3}\,b
\left(\frac{0.2\,\mathrm h^{-1}}{\gamma}\right)
\epsilon^{-2}.
\end{equation}
The reference value $\gamma=0.2\,\mathrm h^{-1}$ is chosen to represent the vicinity of the predicted p53 transmission threshold; it is not inferred from these abundance measurements.

For illustration, at $\epsilon=1$ and $\gamma=0.2\,\mathrm h^{-1}$, a mean burst size $b=10$ gives
\begin{equation}
\Rnoise\simeq
\begin{cases}
0.0182,&\text{mouse fibroblast abundance},\\
0.0505,&\text{HeLa abundance}.
\end{cases}
\end{equation}
These are illustrative parameter choices, since the abundance data alone do not determine $b$.

For an additional turnover benchmark, Schwanh\"ausser \emph{et al.} reported a median protein half-life of $46\,\mathrm h$ \cite{schwanhausser2011global}. Using $\gamma=\ln2/(46\,\mathrm h)$ and $\bar n=5.0\times10^4$ gives a production-flux estimate
\begin{equation}
\gamma\bar n\simeq753.4\,\mathrm h^{-1},
\end{equation}
and hence
\begin{equation}
\bar\rho\simeq\frac{753.4}{b}\,\mathrm h^{-1},
\qquad
\Rnoise\simeq0.0241\,b\,\epsilon^{-2}.
\end{equation}
This calculation combines separate proteome-level medians and should be interpreted as an order-of-magnitude benchmark. If the quoted half-life describes active degradation alone, growth dilution must additionally be included in the effective concentration-loss rate $\gamma$.

\subsection{Effect on the transmission threshold}
For fixed upstream parameters, the critical degradation rate satisfies
\begin{equation}
\gamma_c(\Rnoise)
=\gamma_c(0)\sqrt{1+\Rnoise}.
\end{equation}
Consequently, when $\Rnoise\ll1$,
\begin{equation}
\frac{\gamma_c(\Rnoise)-\gamma_c(0)}
{\gamma_c(0)}
=\frac{\Rnoise}{2}+O(\Rnoise^2).
\end{equation}
The illustrative values $0.0182$ and $0.0505$ above increase the threshold by approximately $0.9\%$ and $2.5\%$, respectively. More generally, a threshold correction below $5\%$ requires
\begin{equation}
\Rnoise<1.05^2-1=0.1025.
\end{equation}
At $\epsilon=1$ and $\gamma=0.2\,\mathrm h^{-1}$, this corresponds approximately to
\begin{equation}
b<
\begin{cases}
56.4,&\text{mouse fibroblast abundance},\\
20.3,&\text{HeLa abundance}.
\end{cases}
\end{equation}

These estimates support neglecting readout noise for sufficiently abundant proteins with sufficiently frequent bursts and appreciable modulation depth. However, proteome-level abundance measurements alone do not establish that the readout noise is negligible for every p53 target.

%%%%%%%%%% references %%%%%%%%%%
\footnotesize\setlength{\bibsep}{5pt}
\bibliography{references}
\bibliographystyle{apsrev4-2}